\documentclass[12pt]{article}
\usepackage[utf8]{inputenc}
\usepackage{tikz}
\usetikzlibrary{arrows}
\usepackage{amssymb,amsthm,verbatim, fontenc,mathtools}

\usepackage{amsmath,amsfonts,graphicx, bm}
\usepackage{vmargin, url}
\usepackage{setspace}
\usepackage[authoryear]{natbib}
\newtheorem{definition}{Definition}[section]
\newtheorem{theorem}{Theorem}[section]

\newtheorem{example}{Example}[section]
\newtheorem{lemma}{Lemma}[section]
\newtheorem{corollary}{Corollary}[section]
\usepackage{color}
\title{A measure of evidence based on the likelihood-ratio statistics}
\author{Alexandre Galv\~ao Patriota\thanks{\footnotesize email: {\tt patriota@ime.usp.br}; fax Brazil:  +55 11 3091-6130}\\
     {\it \footnotesize Departamento de Estat\'istica, IME,
     Universidade de S\~ao Paulo}
     \vspace{-0.4cm}\\
     {\it \footnotesize Rua do Mat\~ao, 1010, S\~ao Paulo/SP, 05508-090, Brazil}
         \\ \\     
}
\date{}
\begin{document}
\maketitle
\begin{abstract}
In this paper, we show that the likelihood-ratio measure (a) is invariant with respect to dominating sigma-finite measures, (b) satisfies logical consequences which are not satisfied by standard $p$-values, (c) respects frequentist properties, i.e., the type I error can be properly controlled, and, under mild regularity conditions, (d) can be used as an upper bound for posterior probabilities. We also discuss a generic  application to test whether the genotype frequencies of a given population are under the Hardy-Weinberg equilibrium, under inbreeding restrictions or under outbreeding restrictions. \\

\end{abstract}

\section{Introduction}

The likelihood ratio (LR) approach has a long history in the statistical literature. \cite{Neyman-Pearson} and \cite{Karlin} demonstrated that the test based on the likelihood ratio statistics is the most powerful for a fixed level of significance under simple hypotheses and monotone LR functions, respectively. \cite{Birkes} studied the relation between generalized LR tests and uniformly most powerful tests. There is a rich literature about this topic, we refer the reader to \cite{Mudholkar}, \cite{Sprott}, \cite{Severini}, \cite{Royall1997}, \cite{Royall2000}, \cite{GiantShenoy2005}, \cite{Blume2008} and \cite{Bickel2012}.

The LR statistic is widely known in the classical statistical inference by its optimal properties and its asymptotic distributions, even under nonstandard regular conditions, that can be employed to make approximate inferences \citep{Chernov1954,Drton}. Typically, $p$-values, test functions and confidence regions can be computed from the asymptotic distribution of this statistic. In this paper, we define the LR measure, discuss some of its random and non-random properties and show that it can be used as an upper bound for the posterior Bayesian probability. An upper bound for the posterior probability can be useful under high dimensional problems were the integration is a cumbersome task. As we shall see, the LR measure is the LR statistic defined as a set function over the subsets of the parameter space.

\begin{figure}[!http]
	\begin{center}
\includegraphics[scale = 0.4]{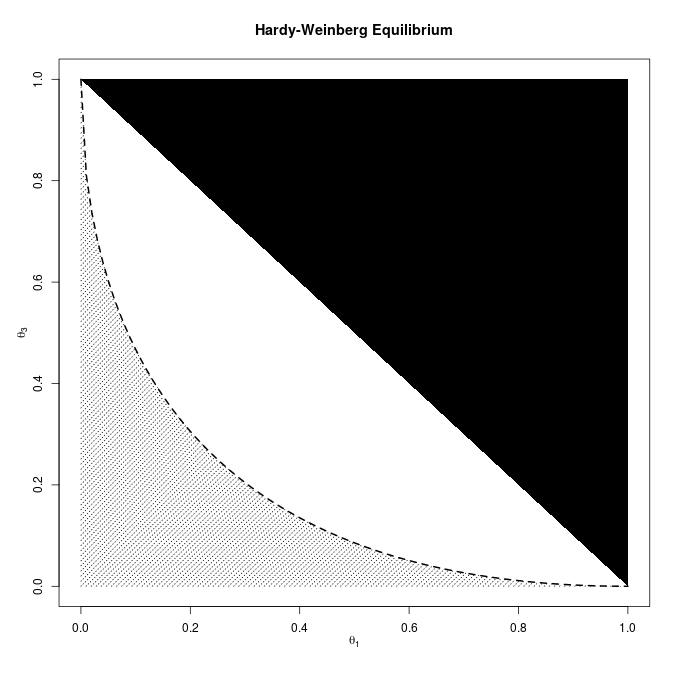}
\caption{The dashed curve represents the Hardy-Weinberg equilibrium. The dotted area below the dashed line represents the inbreeding condition and the white area above the dashed line represents the outbreeding condition. }\label{fig:0}
\end{center}
\end{figure}

This LR-measure can be easily applied to verify whether the genotype frequencies of a given population is under the Hardy-Weinberg Equilibrium (HWE), that is, when those frequencies remain constant from generation to generation. Typically there are three situations for an autosomal biallelic marker: 1) the population is under the HWE; 2) the population is undergoing a regular system of `inbreeding' (when relatives produce offspring); and 3) the population is undergoing a regular system of `outbreeding' (when genetically different individuals produce offspring). In order see these three conditions in terms of the population probabilities, let $AA$, $Aa$ and $aa$ be the possible genotypes and $\theta_1$, $\theta_2$ and $\theta_3$ their respective population probabilities, where $\theta_1+\theta_2+\theta_3 = 1$. From this restriction, we can express the parameter space by using only two parameters, e.g., $\theta_1$ and $\theta_3$ as showed in Figure \ref{fig:0}.

Figure \ref{fig:0} illustrates the HWE in the dashed curve, the inbreeding condition in the dotted area below the dashed line and the outbreeding condition in the white area above the dashed line. Let $y_1$, $y_2$ and $y_3$ be the observed frequencies for the respective genotypes, where $y_1 + y_2 + y_3 = m$, with $m\in \mathbb{N}$. Based on the observed frequencies, one should be able to measure the likelihood the studied population is under the HWE. The farther the point of maximum likelihood $(\frac{y_1}{m}, \frac{y_3}{m})$ is from the dashed curve, the more implausible would be Hardy-Weinberg Equilibrium. The degree of implausibility can be measured by usual frequentist measures such as p-values and confidence regions \citep[see, for instance,][]{Emigh,GraffelmanWeir2016} and Bayesian procedures \citep{Lindley1988,Shoemaker, Puigetal2017}. In this paper, we develop some theoretical results for the likelihood ratio measure and apply them  to the HWE problem in Section \ref{App}.

The rest of this paper is organized as follows. Section \ref{LR-Section} introduces the likelihood-ratio measure, some of its properties and two examples. Section \ref{Bayes-A} provides a comparison with the Bayesian approach. Finally, Section \ref{App} presents an application to the Hardy-Weinberg Equilibrium. 

\section{The likelihood ratio measure}\label{LR-Section}
In order to avoid ambiguity, we adopt the set-measure theory to introduce the likelihood function. Throughout this paper, the statistical model is the triplet $(\mathcal{X}, \mathcal{F}, \mathcal{P})$, where $\mathcal{X}\subseteq \mathbb{R}^n$ is the sample space, $\mathcal{F}$ is a (Borel) sigma-algebra of subsets of $\mathcal{X}$ and $\mathcal{P} = \{P_\theta: \ \theta \in \Theta\}$ is a family of probability measures on $(\mathcal{X}, \mathcal{F})$ that possibly explain the observed data, where $\Theta$ is an arbitrary index set. Each measure $P_\theta \in \mathcal{P}$ is dominated by a sigma-finite measure $\mu$, which we represent by $\mathcal{P}\ll \mu$, that is, $\mu(A) = 0 \Rightarrow P_\theta(A) = 0$ for all $P_\theta \in \mathcal{P}$ and $A \in \mathcal{F}$. If $\Theta \subseteq \mathbb{R}^k$, with $k \in \mathbb{N}$, then the model is said to be parametric; otherwise, the model is said to be nonparametric.  

\begin{definition} The likelihood function is defined to be one version of the Radon-Nikodym derivative 
\[L_\mu(\theta, \cdot) \in \frac{dP_\theta}{d \mu} \quad \mbox{such that} \quad P_\theta(A) = \int_A L(\theta,x) d\mu(x),\] 
where $A \in \mathcal{F}$, $P_\theta \in \mathcal{P}$ and $\frac{dP_\theta}{d \mu}$ is a set containing all versions that are equal except for a set of $\mu$-measure zero.
\end{definition}
The likelihood function is a non-negative function that connects each  probability measure $P_\theta \in \mathcal{P}$ with an observable data $x \in \mathcal{X}$ through a version of the Radon-Nikodym derivative with respect to a sigma-finite measure $\mu$. According to the law of likelihood, the larger is the value of $L_\mu(\theta, x)$, the better is the agreement between the probability $P_\theta$ and the observed data \citep[][p.~71]{Hacking}. In the discrete case, $\mu$ is the counting measure and the likelihood function is $L_\mu(\theta,x) = P_\theta(X = x)$, then $L_\mu(\cdot, x)$ ranks the probabilities in $\mathcal{P}$ from those that make $x$ most probable to those that make $x$ less probable. In the continuous case, $\mu$ is the Lebesgue measure, it can be interpreted as the instantaneous rate of change of $P_\theta$ with respect to $\mu$ at $x$, then $L_\mu(\cdot, x)$  ranks the probabilities in $\mathcal{P}$ from those with the high instantaneous rate of change at $x$ to those with small instantaneous rate of change at $x$. We index the likelihood function by the sigma-finite dominating measure $\mu$ because different sigma-finite dominating measures may generate different likelihood functions.

Let $\Upsilon = \{\mu: \mathcal{P} \ll \mu \}$ be the set of all sigma-finite dominating measures. Throughout this paper, the likelihood function $L_\mu(\theta, \cdot)$ must satisfy
\[
0<  \sup_{\theta \in \Theta} L_\mu(\theta, x) < \infty,  \quad \forall   x \in \mathcal{X}, \ \forall \mu\in \Upsilon. \tag{C$_1$}\label{Cond-1}
\]  Condition (\ref{Cond-1}) guarantees the validity of some properties discussed in this paper. If this supremum is zero for some $x \in \mathcal{X}$, then the family $\mathcal{P}$ in the statistical model was not well specified and should be replaced by an appropriated family. In what follows, we define the LR function and the LR measure.

\begin{definition}
The likelihood ratio function with respect to $\mu$ is 
\begin{equation}\label{LR}
	\lambda_\mu(\theta, x) = \frac{L_\mu(\theta, x)}{\sup\limits_{\theta' \in \Theta}L_\mu(\theta', x)},
\end{equation}
for $\theta \in \Theta$ and $x \in \mathcal{X}$. 
\end{definition}

Theorem below guarantees that the likelihood ratio function does not depend on a specific dominating sigma-finite measure $\mu$.

\begin{theorem}\label{Prop} Let $\mu_1, \mu_2 \in \Upsilon$. Under Condition (\ref{Cond-1}), there exists $A \in \mathcal{F}$ such that $P_\theta(A) = 1$, $\forall \theta \in \Theta$, and
\[
\lambda_{\mu_1}(\theta, x) = \lambda_{\mu_2}(\theta, x) , \quad \forall x \in A, \ \forall \theta \in \Theta.
\] 
\end{theorem}
\begin{proof} Let $L_{\mu_1}$ and $L_{\mu_2}$ be two likelihood functions based on $\mu_1$ and $\mu_2$, respectively, where $\mu_1,\mu_2 \in \Upsilon$. Therefore, according to the The Likelihood Proportionality Theorem derived by \cite{Goncalves-Franklin}, there must exist a measurable function $h$ and a set $A$ such that $P_\theta(A) = 1$, for all $\theta \in \Theta$, and
 \[L_{\mu_1}(\theta, x) = h(x) L_{\mu_2}(\theta, x), \quad \forall  \theta \in \Theta, \ \forall x \in A.\]
Then, by Condition (\ref{Cond-1}), 
 \[
\lambda_{\mu_1}(\theta, x) = \frac{L_{\mu_1}(\theta, x)}{\sup\limits_{\theta' \in \Theta}L_{\mu_1}(\theta', x)}  = \frac{L_{\mu_2}(\theta, x)}{\sup\limits_{\theta' \in \Theta}L_{\mu_2}(\theta', x)} = \lambda_{\mu_2}(\theta, x),
 \] for all $\theta \in \Theta$ and $x \in A.$
\end{proof}

\begin{corollary}\label{Continuous}  Assume valid Condition (\ref{Cond-1}),and let $A$ be the set defined in Theorem \ref{Prop}, where  $x \in A$. If there exists $\mu^* \in \Upsilon$ such that a version $L_{\mu^*}(\cdot, x)$ is continuous on $\Theta$, then $\lambda_\mu(\cdot, x)$ is continuous on $\Theta$ for all $\mu \in \Upsilon$.
\end{corollary}
\begin{proof} It is a direct application of Theorem \ref{Prop}.
\end{proof}

As the likelihood ratio function is well-defined and does not depend on any specific dominating measure, next we define the likelihood measure without indexing by $\mu$.

\begin{definition} Let $A$ be the set defined in Theorem \ref{Prop}. For each fixed $x \in A$, let $\nu_x: 2^{\Theta} \to [0,1]$ be a set function such that
\begin{equation}\label{Set-function}
	\nu_x(\Theta_1) = \sup_{\theta \in \Theta_1}\lambda_\mu(\theta, x) \quad \mbox{and}\quad \nu_x(\varnothing) = 0,
\end{equation} where $\Theta_1\subseteq \Theta$ is a non-empty set and $\mu \in \Upsilon$ is any sigma-finite dominating measure. We say that $\nu_x$ it the likelihood-ratio measure (LR-measure).
\end{definition}

The following properties of $\nu_x$ are derived straightforwardly from its definition. Let $\Theta_1, \Theta_2 \subseteq \Theta$, then
\begin{itemize}
\item[$\mathbb{P}_1$.] $\nu_x(\Theta) = 1$ and $\nu_x(\varnothing) = 0$;
\item[$\mathbb{P}_2$.] $\nu_x(\Theta_1\cup \Theta_2) = \max\{\nu_x(\Theta_1), \nu_x(\Theta_2)\}$;
\item[$\mathbb{P}_3$.] if $\Theta_1$ is non-empty, then $\nu_x(\Theta_1) = \sup\limits_{\theta \in \Theta_1} \nu_x(\{\theta\})$;
\item[$\mathbb{P}_4$.] if $\Theta_1 \subseteq \Theta_2$, then $\nu_x(\Theta_1) \leq \nu_x(\Theta_2)$. 
\end{itemize}

Property $\mathbb{P}_4$ is known as the entailment condition or monotonicity. Properties $\mathbb{P}_1 -\mathbb{P}_4$ indicate that $\nu_x$ is possibility measure rather than a probability measure, which has a non-additive property; see \cite{Denneberg} for more details on non-additive measures and \cite{Zadeh1978} for details on possibility measures.

In order to interpret a $\nu_x$-value, assume that there exists at least one $\theta \in \Theta$ such that $\nu_x(\{\theta\}) = 1$, then:
\begin{quote}
\emph{the quantity $\nu_x(\Theta_0)=\nu_0$ indicates that the highest likelihood produced by the elements of $\Theta_0$ does not exceed $\nu_0\cdot 100\%$ of the largest likelihood.}
\end{quote} 
A small value of $\nu_x(\Theta_0)$ indicates that all cases in $\Theta_0$ produce small likelihoods compared to the largest likelihood (if it exists). Also, in this context, as the largest likelihood occurs in $\Theta$, the complement $\Theta_0^c$, where $\Theta_0^c = \Theta - \Theta_0$, must have full possibility, i.e., $\nu_x(\Theta_0^c)=1$. In other words, the set $\Theta_0$  must be considered ``inconsistent with the observed data $x$'' since the observed data are providing evidence against all element of $\Theta_0$ and no evidence against its complement. It does not mean however that the observed data are providing evidence in favor of its complement $\Theta_0^c$ because other models not listed in $\mathcal{P}$ could be fully possible. This feature suggests that the LR measure may be used as a degree of inconsistency between models and the observed data as has been advocated by \cite{Mudholkar}, \cite{Sprott}, \cite{Severini}, \cite{Royall1997}, \cite{Royall2000}, \cite{GiantShenoy2005}, \cite{Blume2008} and \cite{Bickel2012}.

Next theorem provides a threshold to control (asymptotically) the type I error at significance level $\alpha \in (0,1)$ for testing a general null hypothesis $H_0: {\theta}\in\ \Theta_0$ under some regularity conditions.

\begin{theorem}\label{Test-Size-General} Let $\Theta_0 \subseteq \Theta$ be nonempty and $X$ be the random sample. Assume that $-2 \log(\nu_X(\Theta_0))$ converges in distribution to $Y\sim F_{\Theta_0}$, as $n \to \infty$, for each ${\theta} \in \Theta_0$ where $F_{\Theta_0}$ is a continuous cumulative distribution. Then,
\[
\lim_{n \to \infty} P_{{\theta}} \bigg(\nu_X(\Theta_0) \leq m(\alpha,\Theta_0)\bigg) = \alpha
\] for each $\alpha \in (0,1)$ and ${\theta} \in \Theta_0$, where $m(\alpha,\Theta_0) = \exp(-0.5 F_{\Theta_0}^{-1}(1-\alpha))$.
\end{theorem}
\begin{proof} Let $\alpha \in (0,1)$ and ${\theta}\in \Theta_0$, 
\begin{align*}
\lim_{n \to \infty} P_{{\theta}} \bigg(\nu_X(\Theta_0) \leq \exp(-0.5 F_{\Theta_0}^{-1}(1-\alpha))\bigg) &= \lim_{n \to \infty} P_{{\theta}} \bigg(-2 \log(\nu_X(\Theta_0)) \geq F_{\Theta_0}^{-1}(1-\alpha)\bigg)\\
& = 1 - F_{\Theta_0}(F_{\Theta_0}^{-1}(1-\alpha))\\
& = \alpha.
\end{align*}
\end{proof}
In Theorem \ref{Test-Size-General}, it is assumed that the asymptotic distribution under the null hypothesis $F_{\Theta_0}$ does not depend on $\theta$, but in some nonstandard cases, where $\Theta_0$ includes the boundary of the parameter space or singularities, it could depend \citep[see][]{vu1997generalization, Drton}. Under the conditions of Theorem \ref{Test-Size-General}, the analyst may set the threshold value at $m(\alpha,\Theta_0)$ to control the type I error at  significance level $\alpha$ for testing $H_0: {\theta} \in \Theta_0$ with the rejection rule: ``reject $H_0$ if $\nu_x(\Theta_0) \leq m(\alpha,\Theta_0)$''. This procedure is equivalent to the one based on the asymptotic p-value
\[
\mbox{p-value}(\Theta_0, x) = 1- F_{\Theta_0}(-2\log(\nu_x(\Theta_0)))
\] with the rejection rule: ``reject $H_0$ if $\mbox{p-value}(\Theta_0, x) \leq \alpha$''. The above rejection rules however generate conflicting conclusions on different null hypotheses, since the threshold value $m(\alpha, \Theta_0)$ varies with $\Theta_0$. Because of this feature, it is not guaranteed that the following logical consequence holds:  if $\tilde{H}_0 \to H_0$, then ``rejection of $H_0$'' $\implies$ ``rejection of $\tilde{H}_0$'' with the same data \citep[see, for instance,][]{Schervish, Patriota, Patriota2017}. As showed by \cite{Schervish} and \cite{Patriota}, the $p$-value does not satisfies property $\mathbb{P}_4$  (the entailment condition or monotonicity), but the LR measure satisfies. Therefore, we can find a threshold for the LR measure that satisfies the expected logical consequence and also controls properly the type I error. The following theorem presents this value by employing the asymptotic distribution of $-2 \log(\nu_X(\{{\theta}\}))$ rather than $-2 \log(\nu_X(\Theta_0))$.
\begin{theorem}\label{Test-Size-General-1} Let $\Theta_0 \subseteq \Theta$ be nonempty and $X$ be the random sample. Assume that $-2 \log(\nu_X(\{{\theta}\}))$ converges in distribution to $Y\sim F_{{\theta}}$, as $n \to \infty$, for each ${\theta} \in \Theta$ where $F_{{\theta}}$ is a continuous cumulative distribution. Then,
\[
\lim_{n \to \infty} P_{{\theta}} \bigg(\nu_X(\Theta_0) \leq m(\alpha)\bigg) \leq \alpha
\] for each $\alpha \in (0,1)$ and ${\theta} \in \Theta_0$, where $m(\alpha) = \exp\bigg(-0.5 \sup\limits_{{\theta}\in \Theta} F_{{\theta}}^{-1}(1-\alpha)\bigg)$.
\end{theorem}
\begin{proof} Notice that \[\bigg\{-2 \log(\nu_X(\Theta_0)) \geq \sup\limits_{{\theta}\in \Theta} F_{{\theta}}^{-1}(1-\alpha)\bigg\} \subseteq \bigg\{-2 \log(\nu_X(\{{\theta}\})) \geq F_{{\theta}}^{-1}(1-\alpha)\bigg\}\] for each $\alpha \in (0,1)$ and ${\theta}\in \Theta_0$. Therefore,
\begin{align*}
\lim_{n \to \infty} P_{{\theta}} \bigg(\nu_X(\Theta_0) \leq \exp(-0.5 \sup\limits_{{\theta}\in \Theta} F_{{\theta}}^{-1}(1-\alpha))\bigg) &= \lim_{n \to \infty} P_{{\theta}} \bigg(-2 \log(\nu_X(\Theta_0)) \geq \sup\limits_{{\theta}\in \Theta}F_{{\theta}}^{-1}(1-\alpha)\bigg)\\
&\leq \lim_{n \to \infty} P_{{\theta}} \bigg(-2 \log(\nu_X(\{{\theta}\})) \geq \sup\limits_{{\theta}\in \Theta}F_{{\theta}}^{-1}(1-\alpha)\bigg)\\
&\leq \lim_{n \to \infty} P_{{\theta}} \bigg(-2 \log(\nu_X(\{{\theta}\})) \geq F_{{\theta}}^{-1}(1-\alpha)\bigg)\\
& = 1 - F_{{\theta}}(F_{{\theta}}^{-1}(1-\alpha))\\
& = \alpha
\end{align*}
and the result follows.
\end{proof}

Therefore, under the conditions of Theorem \ref{Test-Size-General-1}, the analyst may set the threshold value at $m(\alpha)$ to control the type I error at significance level $\alpha$ for testing $H_0: {\theta} \in \Theta_0$ with the rejection rule: ``reject $H_0$ if $\nu_x(\Theta_0) \leq m(\alpha)$''. As this threshold value does not depend on the null parameter space $\Theta_0$, the latter rejection rule does not generate conflicting conclusions on different null hypotheses. Because of this feature, the expected logical consequence holds:  if $\tilde{H}_0 \to H_0$, then ``rejection of $H_0$'' $\implies$ ``rejection of $\tilde{H}_0$'' with the same data. Under standard regularity conditions, the asymptotic distribution $F_{{\theta}}$ is a chisquared cumulative distribution with $k = \mbox{dim}(\Theta)$ degrees-of-freedom and therefore it does not depend on ${\theta}$.

The LR measure can be rewritten as indicated in the next lemma. This characterization will help to illustrate the example presented in Section \ref{App}.

\begin{lemma}\label{Def2} Let $A$ be the set defined in Theorem \ref{Prop}. Then, for each fixed $x \in A$ and each $\theta \in \Theta$,  
\[\nu_x(\{\theta\}) = \sup\{\alpha \in [0,1]: \ \Lambda_\alpha(x) \cap\{\theta\} \neq \varnothing\},
\] where $\Lambda_\alpha(x) = \{\theta \in \Theta: \lambda_\mu(\theta, x)\geq \alpha\}$ and $\mu\in \Upsilon$.
\end{lemma}
\begin{proof}
Since for each $x \in A$ and $\theta \in \Theta$, we have that
\[\big\{\alpha \in [0,1]: \Lambda_\alpha(x) \cap \{\theta\}\neq \varnothing\big\} =	\big\{\alpha \in [0,1]: \lambda_\mu(\theta, x)\geq \alpha\} = \big[0, \lambda_\mu(\theta, x)\big],\] for each $\mu\in \Upsilon$. The proof is completed.
\end{proof}

The set $\Lambda_{\nu_0}(x)$, where $\nu_0 = \nu_x(\Theta_0)$ highlights the distance of $\Theta_0$ from the set $\widehat{\Theta}_x = \{\theta \in \Theta: \ \nu_x(\{\theta\})=1\}$ of points that yields maximum likelihoods (if it is nonempty). The more distant is the border of $\Lambda_{\nu_0}$ from its center, the more implausible is $\Theta_0$ according to the likelihood function.

\subsection{Two illustrative examples}\label{Examples}
We consider two simple examples from Binomial and Poisson distributions in Figures \ref{Bin-fig} and \ref{Pois-fig}, respectively.

\begin{example} (Binomial) Let $\mathcal{X} = \{0,1,\ldots, 8\}$ and $X$ be a binomial random variable such that $P_\theta(X = x) = {8 \choose x}\theta^x(1-\theta)^{8-x}$, for $x \in \mathcal{X}$ and $\theta \in (0,1)$. Figure \ref{Bin-fig} shows the $\nu_x$-values for $\Theta_0 = [0.4,0.6]$ and all observable values of $x \in \mathcal{X}$. The LR-measure is 
\[\nu_x(\{\theta\}) = \left\{
\begin{array}{ll}
	\dfrac{\theta^x(1-\theta)^{8-x}}{(\frac{x}{8})^x(1-\frac{x}{8})^{8-x}}, & \mbox{ if } x \in \{1,\ldots,7\}\\
	\theta^x(1-\theta)^{8-x}, & \mbox{ if } x\in \{0,8\}.
\end{array}\right.\] 

As can be seen in Figure \ref{Bin-fig}(e), $\nu_4(\Theta_0) = 1$ and $\nu_4(\Theta_0^c) = 0.85$, that is, when $x=4$ the set $\Theta_0$ contains the points that generates the highest likelihood. The likelihoods produced by the elements in $\Theta_0^c = (0,0.4) \cup (0.6,1)$ attain at most 85\% of the maximum likelihood.

In Figure \ref{Bin-fig}(a) and (i), $\nu_x(\Theta_0) = 0.02$ and $\nu_x(\Theta_0^c) = 1$ for either $x=0$ and $x=8$. That is, when $x\in \{0, 8\}$, the set $\Theta_0^c$ contains the elements that generates the highest likelihood and the likelihoods produced by the elements of $\Theta_0$ reach at most 2\% of the maximum likelihood. Observe that the observed values $x\in\{0,8\}$ are the ones that provide the most evidence against $\Theta_0$. The interpretations for the other cases are similar.
\begin{figure}[!htp]
\begin{center}
\includegraphics[scale=0.6]{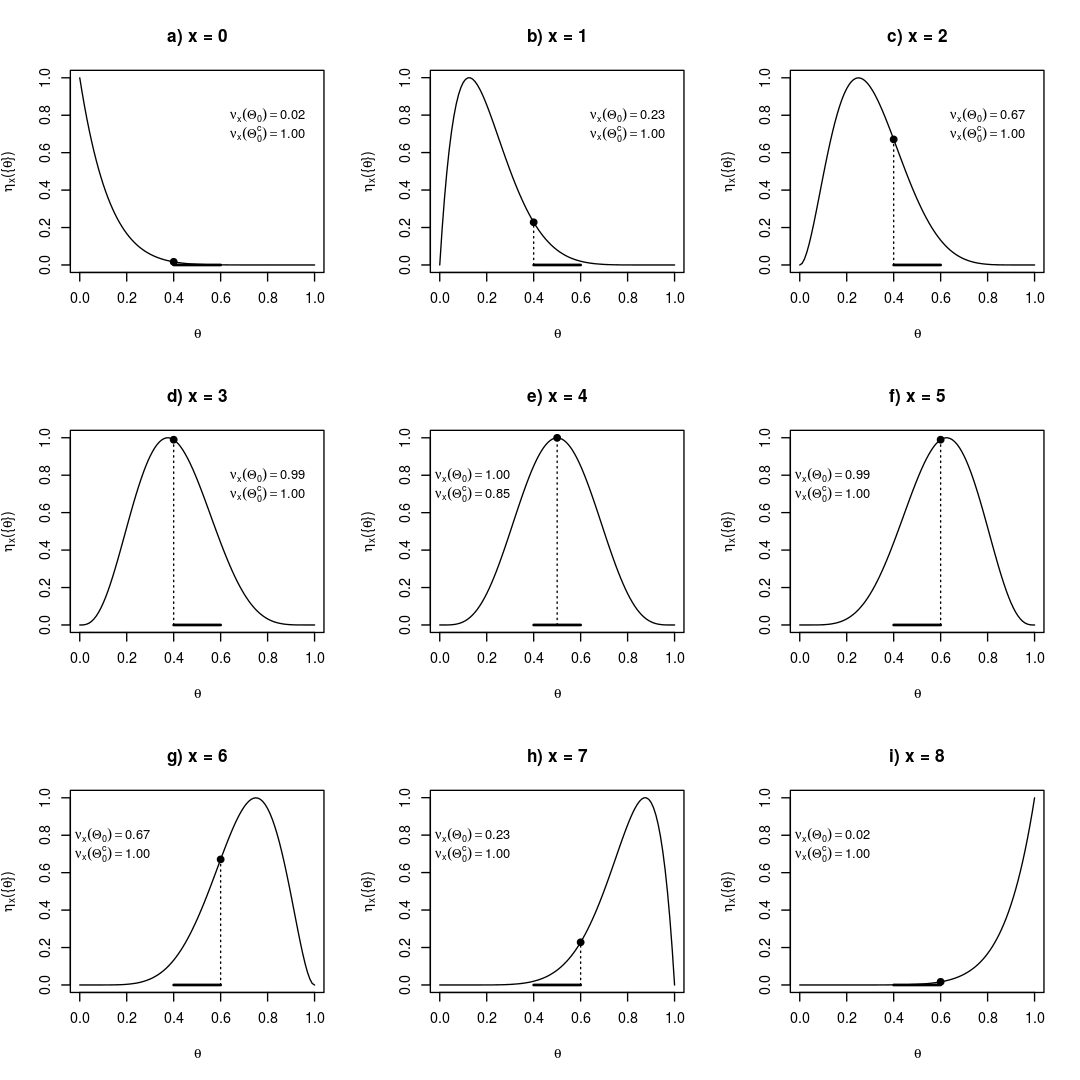}
\end{center}
\caption{Normalized likelihood curve for all observable values from a binomial random variable with $n=8$. The horizontal full segment represents the set $\Theta_0 = [0.4,0.6]$. The big dot represents the value of $\nu_x(\Theta_0)$ on the curve.}\label{Bin-fig}
\end{figure}
\end{example}

\begin{example} (Poisson) Let $\mathcal{X} = \{0,1,\ldots\}$ and $X$ be a Poisson random variable, i.e.,  $P_\theta(X = x) = \exp(-\theta)\theta^x/x!$, for $x \in \mathcal{X}$ and $\theta \in (0,\infty)$. Figure \ref{Pois-fig} depicts the $\nu_x$-values for the first eight observable values for  $\Theta_0 = (0,3]$. The LR-measure is
\[\nu_x(\{\theta\}) = \left\{
\begin{array}{ll}
\exp(x-\theta)\big(\frac{\theta}{x}\big)^x, & \mbox{ if } x \in \{1,2,\ldots\}\\
\exp(-\theta), & \mbox{ if } x = 0.
\end{array}\right.\]
Figure \ref{Pois-fig}(a) shows that $\nu_0(\Theta_0) = 1.00$ and $\nu_0(\Theta_0^c) = 0.05$. Notice that, when $x=0$, there is no element in $\Theta$ that yields the highest likelihood, but the least upper bound of $\mu_0(\{\theta\})$ is $\sup\limits_{\theta >0} \exp(-\theta)= 1$. The elements in $\Theta_0^c = (3, \infty)$ attain at most 5\% of the least upper bound for all likelihoods in this model. Figure \ref{Pois-fig}(i) shows that $\nu_8(\Theta_0) = 0.06$ and $\nu_8(\Theta_0^c) = 1.00$, that is, when $x=8$, the element that yields the largest likelihood is in $\Theta_0^c$ while the elements in $\Theta_0$ attain at most 6\% of the largest likelihood. The interpretation for the other cases is similar.
\begin{figure}[!htp]
\begin{center}
\includegraphics[scale=0.6]{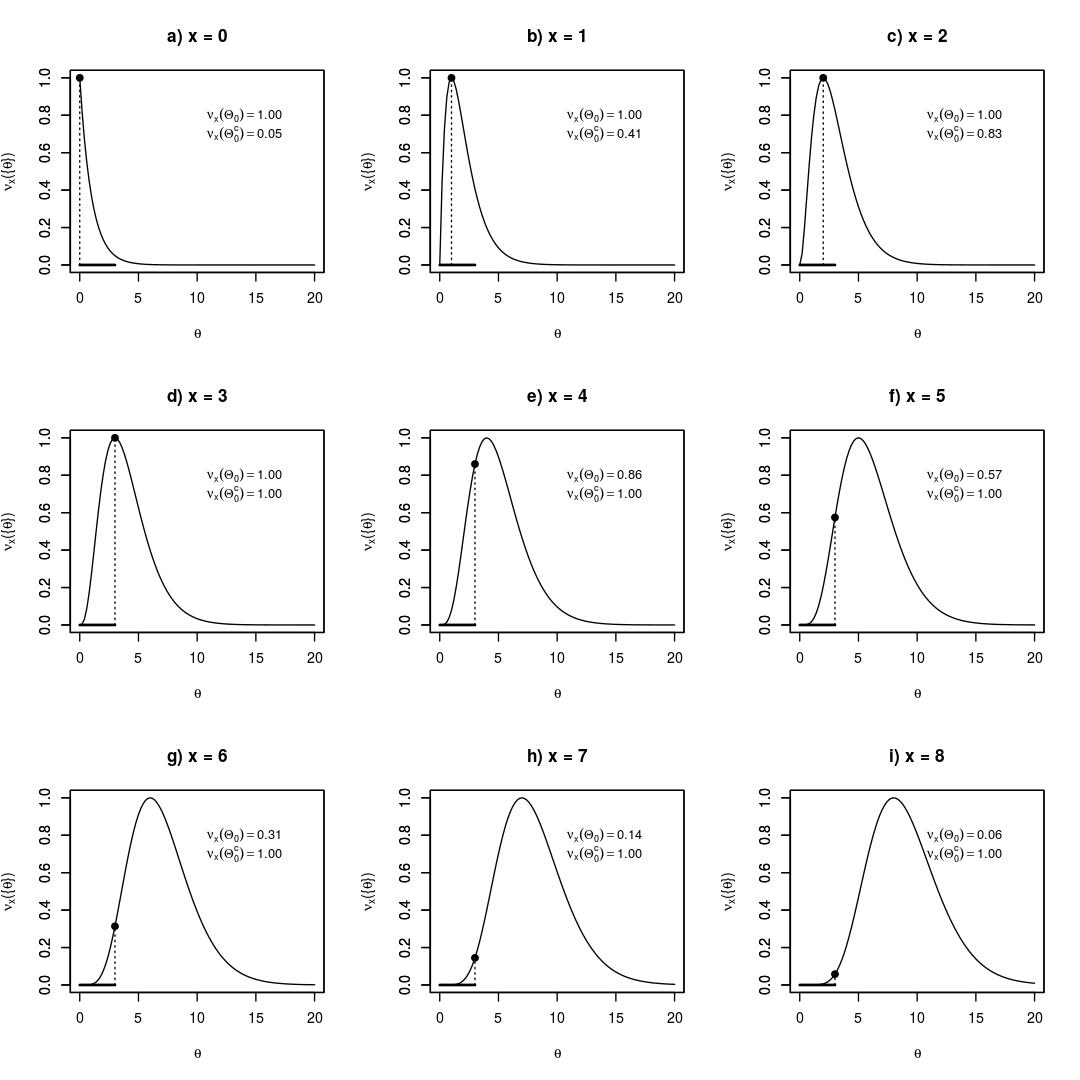}
\end{center}
\caption{Normalized likelihood curve for the first eight observable values of a Poisson random variable. The horizontal full segment represents the set $\Theta_0 = (0,3]$. The big dot represents the value of $\nu_x(\Theta_0)$ on the curve.}\label{Pois-fig}
\end{figure}
\end{example}

In both examples, as the standard regularity conditions hold, the asymptotic distribution of $-2\log(\nu_X(\{\theta\}))$ is a chisquared distribution with one degree-of-freedom. In this context, the threshold derived in Theorem \ref{Test-Size-General-1} is $m(\alpha) = \exp(-0.5 F_{\chi_1^2}^{-1}(1-\alpha))$, where $F_{\chi_1^2}$ is the cumulative chisquared distribution with one degree-of-freedom. In order to reject at 5\% significance level and to maintain monotonicity (the entailment condition or the logical consequences), we need to set the threshold value at $m(0.05) = 0.1465$.





\subsection{Comparison with the Bayesian approach}\label{Bayes-A}

In the Bayesian framework, a prior probability $\pi$, dominated by a sigma-finite measure $\xi$, is defined over the Borel sigma-field of $\Theta$ from which a posterior is attained through \[h_x(\theta) = \frac{L_\mu(\theta, x)h(\theta)}{m_\mu(x)},\] where
$m_\mu(x)= \int_{\Theta} L_\mu(\theta,x)d\pi(\theta)$ is the normalizing constant and $h \in \frac{d\pi}{d\xi}$ is one version of the Radon-Nikodym derivative of $\pi$ with respect to $\xi$.  The posterior probability of $\Theta_0$ given the observed {data} $x$ is defined by the Lebesgue integral
\[
\pi_x(\Theta_0) = \int_{\Theta_0} h_x(\theta)d\xi(\theta).
\]   If $\pi_x(\Theta_0)$ is very small, then $\Theta_0$ is improbable according to $\pi_x$. In this probabilistic framework, the value $\pi_x(\Theta_0^c)$ is completely determined by $\pi_x(\Theta_0)$, since  $\pi_x(\Theta_0^c) = 1 - \pi_x(\Theta_0)$. Then, the following consequence is straightforward
\[0<\pi_x(\Theta_0)< \pi_x(\Theta_1)<1 \Longleftrightarrow 0<\pi_x(\Theta_1^c)<\pi_x(\Theta_0^c) < 1.\] By properties $\mathbb{P}_1-\mathbb{P}_4$, this consequence does not hold for the LR-measure $\nu_x$, since  $\nu_x(\Theta_0) <1 \Rightarrow \nu_x(\Theta_0^c) = 1$ which justifies instead the following relation
\begin{equation}\label{Rel2}
0<\nu_x(\Theta_0) < \nu_x(\Theta_1)<1 \Longrightarrow \nu_x(\Theta_0^c)= \nu_x(\Theta_1^c)=1.
\end{equation} That is, an increasing in the $\nu_x$-values from $\Theta_0$ to $\Theta_1$ does not necessarily imply a decreasing in the $\nu_x$-values from $\Theta_0^c$ to $\Theta_1^c$. They may remain constant. Although $\nu_x$ and $\pi_x$ have different properties,  they are related in the following lemma. 

\begin{lemma}\label{dominate} Let $\pi$ be a proper prior probability dominated by a sigma-finite measure and defined over a measurable list of subsets of $\Theta$. Assume valid Condition (\ref{Cond-1}) and let $A$ the set defined in Theorem \ref{Prop}, then, for any $\pi$-measurable $\Theta_0$ such that $\pi(\Theta_0)> 0$, 
\[
\nu_x(\Theta_0) \geq  \frac{\pi_x(\Theta_0)m_\mu(x)}{\pi(\Theta_0)c_\mu(x)},
\]
for any $x\in A$, where $c_\mu(x) = \sup\limits_{\theta \in \Theta}L_\mu(\theta,x)$ and $\mu \in \Upsilon$.
\end{lemma}
\begin{proof} As $0 < c_\mu(x)<\infty$ for each $x \in A$, the posterior probability can be written as
\begin{equation}\label{zero}
	\pi_x(\Theta_0) = \frac{c_\mu(x)}{m_\mu(x)}\int_{\Theta_0} \lambda_\mu( \theta,x)d\pi(\theta) \leq  
\frac{c_\mu(x)}{m_\mu(x)}\pi(\Theta_0)\sup_{\theta \in \Theta_0} \lambda_\mu(\theta,x),
\end{equation} therefore the result follows
\[
\nu_x(\Theta_0) \geq  \frac{\pi_x(\Theta_0)m_\mu(x)}{\pi(\Theta_0)c_\mu(x)}.
\]

\end{proof}

The set function $\nu_x$ is defined over the power set of $\Theta$, while the posterior distribution is defined over the family of measurable subsets (in the Lebesgue sense) of $\Theta$. Thus, there may exist subsets of $\Theta$ that are not measurable by $\pi_x$ but are computable by $\nu_x$. Moreover, from Equation (\ref{zero}), $\nu_x(\Theta_0) = 0$ implies that $\pi_x(\Theta_0) = 0$ but the converse is not true, for there exist zero prior probabilities for $\Theta_0$. In this sense, $\nu_x$ dominates $\pi_x$. This feature is in agreement with the expected reasoning between possibility and probability:
\begin{quote}
``a high degree of possibility does not imply a high degree of probability, nor does a low degree of probability imply a low degree of possibility. However, if an event is impossible, it is bound to be improbable.'' \citep{Zadeh1978}
\end{quote}

Lemma \ref{dominate} provides upper bound for the posterior probability of $\Theta_0$. The following corollary is straightforwardly attained from Lemma \ref{dominate}.
\begin{corollary}\label{Consistency}
Assume valid Lemma \ref{dominate}'s assumptions. Let $\Theta_0$ be a $\pi$-measurable set. Then,

\begin{itemize}
\item[1.] $\pi(\Theta_0) \leq \frac{m_\mu(x)}{c_\mu(x)} \implies 
\pi_x(\Theta_0) \leq \nu_x(\Theta_0)$,
\item[2.] $\nu_x(\Theta_0) \leq \frac{m_\mu(x)}{c_\mu(x)} \implies 
\pi_x(\Theta_0) \leq \pi(\Theta_0)$.
\end{itemize}
\end{corollary}
The extra condition of the Corollary \ref{Consistency} is not vacuous, since $\frac{m_\mu(x)}{c_\mu(x)} \leq 1$ by the following
\[
m_\mu(x)= \int_{\Theta} L_\mu(\theta,x)d\pi(\theta) \leq \sup_{\theta \in \Theta} L_\mu(\theta, x) \pi(\Theta) = c_\mu(x).
\] Corollary \ref{Consistency} holds for measurable subsets of $\Theta$ with low prior probabilities relative to $m_\mu(x)/c_\mu(x)$. This condition is sufficient but not necessary, since the set $\Theta_0$ can have a high probability prior and a relatively low maximum likelihood, then its posterior probability might be much smaller than the prior probability. 


It is worth noting that only optimization procedures are required in the computation of  $\nu_x(\Theta_0)$, no integration is required. Computing a posterior probability can be a cumbersome task in some high dimensional problems, therefore, an upper bound of the posterior probability can be attained by Lemma \ref{dominate} and Corollary \ref{Consistency} in terms of $\nu_x$. That is, if the $\nu_x$-value is small for an event, then the posterior probability must be even smaller (under the specified conditions), unless the prior is too distant from the region where likelihood is concentrated.

\section{Application to the Hardy-Weinberg Equilibrium}\label{App}

In this section, we apply the above results to analyze genotype frequencies in a population. We investigate three scenarios, namely: 1) the population is under the Hardy Weinberg equilibrium; 2) the population is undergoing a regular system of `inbreeding' (when relatives produce offspring); and 3) the population is undergoing a regular system of `outbreeding' (when very genetically different individuals produce offspring). These scenarios are formalized mathematically in terms of sets the sequel.

Let $AA$, $Aa$ and $aa$ be the possible genotypes and $\theta_1$, $\theta_2$ and $\theta_3$ their respective population frequencies, where $\theta_1+\theta_2+\theta_3 = 1$. The parameter vector is $\theta = (\theta_1, \theta_2, \theta_3)$. Denote $y_1$, $y_2$ and $y_3$ the observed frequencies of the genotypes $AA$, $Aa$ and $aa$, respectively, where $y_1+y_2+y_3 = m$.

The statistical model is then defined in the following. Let $(\mathcal{X}, \mathcal{F}, \mathcal{P})$ be a parametric statistical model, where \[\mathcal{X} = \{(y_1,y_2,y_3) \in \mathbb{N}^3: \ y_1 + y_2 + y_3 = m\},\] $m\in \mathbb{N}$ is a fixed value, $\mathcal{F} = 2^{\mathcal{X}}$ and each $P_\theta \in \mathcal{P}$ is defined by
\[
P_\theta(A) = \sum_{(y_1,y_2,y_3)  \in A} \frac{m!}{y_1!y_2!y_3!} \theta_1^{y_1} \theta_2^{y_2}\theta_3^{y_3},
\] where $A \in \mathcal{F}$ is a nonempty measurable set and the parameter space is \[\Theta = \{\theta \in [0,1]^3: \ \theta_1 + \theta_2 + \theta_3 = 1\}.\] For the observed sample $x = (y_1,y_2,y_3)$, the likelihood function and the likelihood ratio statistics are, respectively, given by
\[
L(x,\theta) = \frac{m!}{y_1!y_2!y_3!} \theta_1^{y_1} \theta_2^{y_2} \theta_3^{y_3}\quad \mbox{and} \quad 
\lambda(x, \theta) = c\theta_1^{y_1}\theta_2^{y_2}\theta_3^{y_3}, \]
where $c = \prod\limits_{i: y_i>0} \bigg(\frac{m}{y_i}\bigg)^{y_i}$.

The population is under the Hardy-Weinberg equilibrium when $\sqrt{\theta_3} = 1- \sqrt{\theta_1}$, for this situation we define the set \[\Theta_{1} = \bigg\{\theta \in \Theta: \ \sqrt{\theta_3} = 1- \sqrt{\theta_1}\bigg\}.\] The population is under an inbreeding pressure when $\sqrt{\theta_3} < 1- \sqrt{\theta_1}$, for this case we define the set \[\Theta_{2} = \bigg\{\theta \in \Theta: \ \sqrt{\theta_3} < 1- \sqrt{\theta_1}\bigg\}.\] Finally, the population is under an outbreeding pressure when $\sqrt{\theta_3} > 1- \sqrt{\theta_1}$, for this last case we define the set 
\[\Theta_{3} = \bigg\{\theta \in \Theta: \ \sqrt{\theta_3} > 1- \sqrt{\theta_1}\bigg\}.\] For more details and discussion on this topic the reader is referred to \cite{Emigh}. The LR-measure for each set is 
\[
\nu_x(\Theta_i) = \sup_{\theta \in \Theta_i} \lambda(\theta,x)
\] for $i=1,2,3$. Notice that
\[
\nu_x(\Theta_1) = c \sup_{\theta \in \Theta_1} \theta_1^{y_1}\theta_2^{y_2}\theta_3^{y_3},
\] where $\theta_3 = (1 - \sqrt{\theta_1})^2$ and $\theta_2 = 1 - \theta_1 -  (1 - \sqrt{\theta_1})^2$, that is, $\lambda(x,\theta)$ restricted to $\Theta_1$ depends only on $\theta_1$. Then, we can rewrite the $\nu_x$-value by
\[
\nu_x(\Theta_1)  = c \sup_{\theta_1 \in [0,1]} f(\theta_1),
\] where $f(z) = z^{y_1}(1 - z - (1-\sqrt{z})^2)^{y_2}(1-\sqrt{z})^{2y_3}$. As $f$ is continuous, $f([0,1])$ is closed and it has a maximum, i.e., $\sup f([0,1]) = \max f([0,1])$. It is possible to show that the maximum value of $f$ is attained at 
\[z = \frac{(m + y_1 -y_3)^2}{4m^2}.\]

Therefore, the $\nu_x$-value for $\Theta_1$ is
\[
	\nu_x(\Theta_1) = \lambda(x, \widetilde{\theta}),
\] where $\widetilde{\theta} = (\widetilde{\theta}_1,\widetilde{\theta}_2,\widetilde{\theta}_3)$, $\widetilde{\theta}_1 = \frac{(m + y_1 -y_3)^2}{4m^2}$, $\widetilde{\theta}_2 = 1 - \widetilde{\theta}_1 -  (1 - \sqrt{\widetilde{\theta}_1})^2$ and  $\widetilde{\theta}_3 = (1 - \sqrt{\widetilde{\theta}_1})^2$. Notice that, \[\mbox{dim}(\Theta_1) = 1, \quad \mbox{dim}(\Theta_2) =\mbox{dim}(\Theta_3) = \mbox{dim}(\Theta) =2,\] also note that for each $0 < \theta_1, \theta_2, \theta_3 < 1$, the function $\log(\lambda(x,\cdot))$ is concave. Therefore, the maximum likelihood estimate $\widehat{\theta}$ satisfies 
\[\widehat{\theta}\in \overline{\Theta_2} \quad \mbox{ or } \quad \widehat{\theta}\in \overline{\Theta_3},\] where $\overline{A}$ is the closure of set $A$. Also, we have the following three situations
\begin{enumerate}
	\item If  $\widehat{\theta}\in \overline{\Theta_2}$, then $\nu_x(\Theta_2)  = 1$ and $\nu_x(\Theta_3) = \nu_x(\Theta_1)$.
	\item If  $\widehat{\theta}\in \overline{\Theta_3}$, then $\nu_x(\Theta_3) = 1$ and $ \nu_x(\Theta_2)  = \nu_x(\Theta_1)$.
	\item If  $\widehat{\theta}\in \overline{\Theta_2}\cap \overline{\Theta_3}$, then $\nu_x(\Theta_i) = 1$, for $i=1,2,3$.
	\end{enumerate}
Hence, in any cases, we only need to compute numerically $\nu_x(\Theta_1)$. The asymptotic distribution of $-2\log(\lambda(X,\theta))$ is a chisquared distribution with two degrees-of-freedom for each $\theta \in \Theta$. In this context, the asymptotic threshold derived in Theorem \ref{Test-Size-General-1} is $m(\alpha) = \exp(-0.5 F_{\chi_2^2}^{-1}(1-\alpha)) = \alpha$, where $F_{\chi_2^2}$ is the cumulative chisquared distribution with two degrees-of-freedom. In order to reject at 5\% significance level and to maintain monotonicity (the entailment condition or the logical consequences), we choose the threshold value at $m(0.05) = 0.05$. 

\begin{figure}[!htp]
	\begin{center}
\includegraphics[scale = 0.42]{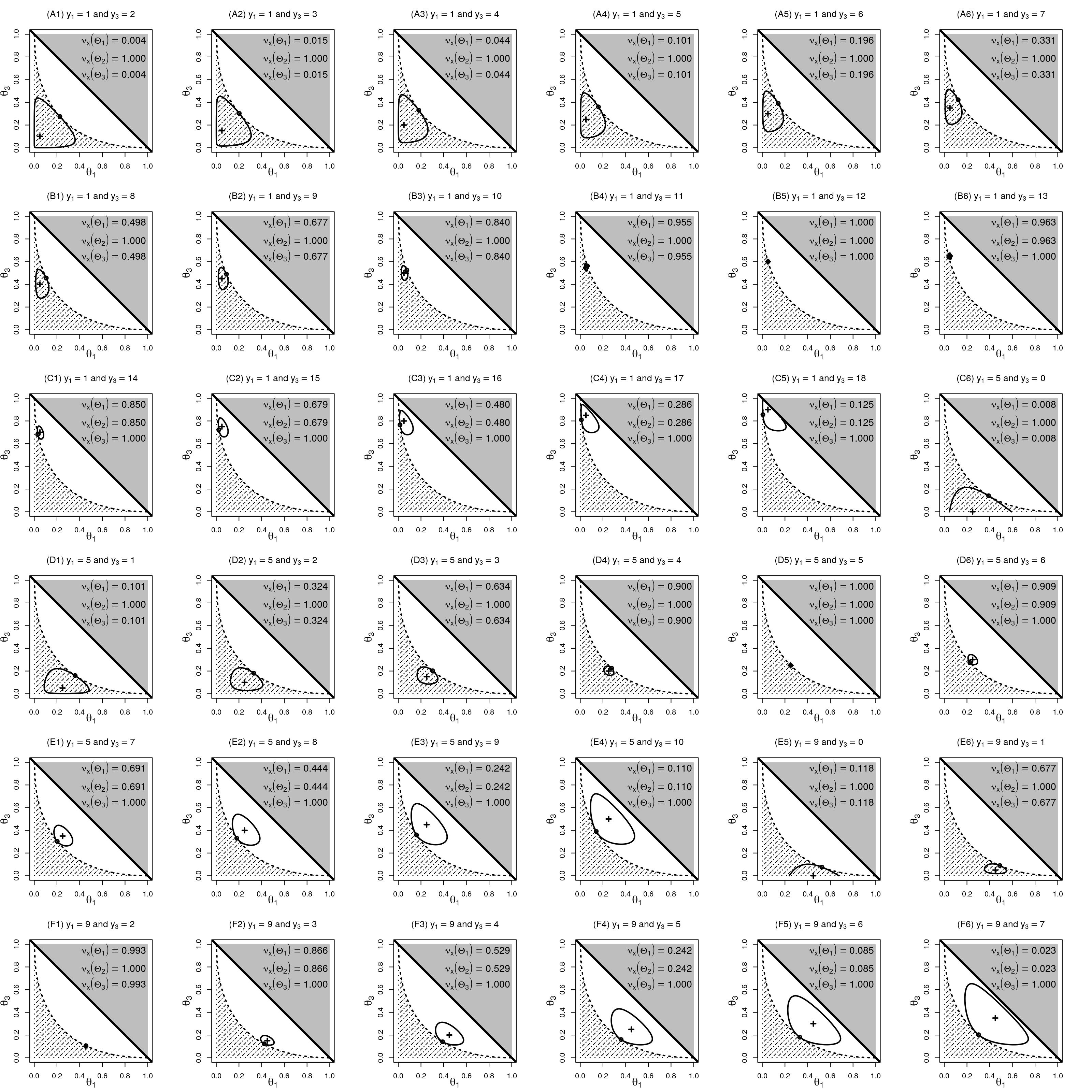}
\caption{The dashed curves represent the Hardy-Weinberg equilibrium ($\Theta_1$). The crosshatched area below the dashed line represents the inbreeding condition ($\Theta_2$) and the white area above the dashed line represents the outbreeding condition ($\Theta_3$). The cross mark is the maximum likelihood estimative for each observed sample $(y_1, y_3)$ and its surrounding full line is the smallest contour for which $\Lambda_\alpha$ has at least one element in common with $\Theta_1$ (dashed line). The associated $s$-values for $\Theta_1$, $\Theta_2$ and $\Theta_3$ are presented.}\label{fig:2}
\end{center}
\end{figure}

In Figure \ref{fig:2}, the simplex formed by points (0,0), (0,1) and (1,0) represents the parameter space $\Theta$ in a reduced form (only with $\theta_1$ and $\theta_3$, for the restriction $\theta_1 + \theta_2 + \theta_3 =1$ and also we consider $m=20$). Each plot presents the three sets $\Theta_1$, $\Theta_2$ and $\Theta_3$ in their respective reduced forms (only with $\theta_1$ and $\theta_3$). The dashed curve illustrates the Hardy-Weinberg equilibrium ($\Theta_1$). The crosshatched area (below the dashed curve) stands for the inbreeding restriction ($\Theta_2$) and the white area (above the dashed curve) stands for the outbreeding restriction ($\Theta_3$). Each plot in Figure \ref{fig:2} refers to a specific observed sample, for instance plot (A1) refers to $y_1=1$ and $y_3=2$, plot (A2) refers to $y_1=1$ and $y_3=3$ and so forth. In each plot, the cross mark indicates the maximum likelihood estimate for $(\theta_1, \theta_3)$ and the full lines surrounding the ML estimates are the smallest contours for which $\Lambda_\alpha$ has at least one element in common with $\Theta_1$. 

On the one hand, in the plots (A1)--(A6), (B1)--(B4), (C6), (D1)--(D4), (E5)--(E6) and (F1) the maximum likelihood lies in $\Theta_2$, that is, it is favoring the inbreeding restriction. Only for (A1)-(A3) and (A6),  this favoring is relevant at 5\% significance level: 

\begin{itemize}
	\item (A1): $y_1=1$, $y_3=2 \ \Rightarrow \ \nu_x(\Theta_3) = 0.004$,
	\item (A2): $y_1=1$, $y_3=3 \ \Rightarrow \ \nu_x(\Theta_3) = 0.015$, 
	\item (A3): $y_1=1$, $y_3=4 \ \Rightarrow \ \nu_x(\Theta_3) = 0.044$,
	\item (C6): $y_1=5$, $y_3=0 \ \Rightarrow \ \nu_x(\Theta_3) = 0.008$.
\end{itemize}		
 
On the other hand, in the plots (B6), (C1)--(C5), (D6), (E1)--(E4) and (F2)--(F6) the maximum likelihood lies in $\Theta_3$, that is, it is favoring the outbreeding restriction. Only for (F6),  this favoring is  relevant at 5\% significance level: 
\begin{itemize}
	\item (F6): $y_1=9$, $y_3=7 \ \Rightarrow \ \nu_x(\Theta_2) = 0.023$.
\end{itemize}		

It is also noteworthy that in two cases the maximum likelihood is in $\overline{\Theta_2}\cap \overline{\Theta_3} = \Theta_1$, namely: (B5) and (D5). In those cases, we have that $\nu_x(\Theta_1) = \nu_x(\Theta_2) = \nu_x(\Theta_3) = 1$ which means no evidence against any hypotheses.

\section{Concluding remarks}\label{concluding}

This paper discussed theoretical properties of the likelihood-ratio measure for random and observed samples. We showed that the LR measure (a) satisfies logical consequences and frequentist principles and (b) it  can be used as upper bounds for posterior probabilities with relative small prior probabilities. One can properly control the type I error in a hypothesis testing procedure without violating logical consequences over the set of null hypotheses. It should be clear that testing different null hypotheses should not be confused with sequential hypothesis testing, since the former does not employ the result of a test to conduct another one. An application to the Hardy-Weinberg equilibrium was presented.

\section{Acknowledgements}
This work received grants from FAPESP--Brazil (2014/25595-0) and CNPq (200115/2015-4). This paper was partially developed in the Department of Biochemistry, Microbiology, and Immunology, University of Ottawa, Canada,  and in Department of Statistics, University of S\~ao Paulo, Brazil.

\end{document}